*Proceeding Paper*

# Forecasting Tangency Portfolios and Investing in the Minimum Euclidean Distance Portfolio to Maximize Out-of-Sample Sharpe Ratios †

Nolan Alexander * and William Scherer

Department of Systems Engineering, University of Virginia, Charlottesville 22903, VA, USA; wts@virginia.edu
* Correspondence: nolan_alex2018@yahoo.com
† Presented at the 9th International Conference on Time Series and Forecasting, Gran Canaria, Spain, 12–14 July 2023.

**Abstract:** We propose a novel model to achieve superior out-of-sample Sharpe ratios. While most research in asset allocation focuses on estimating the return vector and covariance matrix, the first component of our novel model instead forecasts the future tangency portfolio, and the second component then determines the optimal investment portfolio. First, to forecast the tangency portfolio, we forecast the efficient frontier by decomposing its functional form, a square root second-order polynomial, into three interpretable coefficients, which can then be used to calculate a forecasted tangency portfolio. These coefficients can be forecasted using vector autoregressions. Second, the model invests in the portfolio on the efficient frontier that is the minimum Euclidean distance from this forecasted tangency portfolio. A motivation for our approach is to address the limitation that the tangency portfolio only maximizes the Sharpe ratio when future returns and covariances are stationary, and can be directly estimated with historical data, which often does not hold in out-of-sample data. Our approach addresses this shortcoming in a novel way by forecasting the tangency portfolio, rather than estimating return and covariance. For empirical testing, we employ two sets of assets that span the market to demonstrate and validate the performance of this novel method.

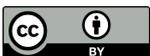

## 1. Introduction

Mean–variance optimization, developed by Markowitz [1], is the foundation to Modern Portfolio Theory (MPT). The model uses quadratic programming to select portfolios that minimize risk, as measured by volatility, while generating a fixed expected return. The model constructs a pareto frontier that consists of these optimal portfolios, known as an efficient frontier. This efficient frontier allows investors to visualize tradeoffs between portfolio risk and return. A natural question that arises is which portfolio on the efficient frontier is best to invest in. In 1964, William Sharpe developed a portfolio metric known as the Sharpe ratio, defined as the ratio of the portfolio return exceeding the risk-free rate and the portfolio's volatility [2]. The portfolio that maximizes the Sharpe ratio when there is a risk-free rate is the one that corresponds with the market. This portfolio intersects with the capital market line (CML) under the Capital Asset Pricing Model (CAPM), which has an intercept at the risk-free rate and a slope defining the Sharpe ratio. This Sharpe-maximizing portfolio is known as the tangency portfolio because the CML is tangent to the efficient frontier as the feasible region is convex.

The tangency portfolio model makes several assumptions that researchers have attempted to relax, such as transaction costs [3]. This paper, however, will only focus on relaxing one: that the future expected asset returns and covariances will be identical to the asset's historical data. This paper will focus on relaxing this assumption, as there often exist significant estimation errors in practice because of nonstationary market behavior. This limitation is significant, as Dickinson [4] found that the misestimation of these parameters causes the portfolio weights to be unstable, and Kan and Zhou [5] found that estimating the population parameters with samples leads to poor performance out-of-sample. There exist multiple models that attempt to relax this assumption already. Some of the most significant are the Black–Litterman model, Covariance Matrix Shrinkage, and Dynamic Conditional Correlation Multivariate GARCH (DCC MV-GARCH).

In 1992, Black and Litterman [6] proposed a closed-form model to provide better posterior estimates of the return vector and covariance matrix using a Bayesian approach. The Black–Litterman model uses two priors: investors' beliefs in the assets and the market capitalized equilibrium view, which weights assets by their market capitalization. The model requires investors to provide several inputs: a matrix of their views on the assets, a vector of the returns in each view, the level-of-unconfidence in each view, and the hyperparameter weight-on-views, which determines how much weight should be put on each prior.

Engle and Sheppard [7] proposed a method to forecast covariance matrices using Generalized Autoregressive Conditional Heteroskedasticity (GARCH), which is a time-series method to estimate volatility (Engle and Sheppard, 2001). DCC MV-GARCH transforms residuals from GARCH models applied to each asset to obtain conditional correlation estimators, which can be used to estimate the covariance matrix.

Ledoit and Wolf [8] proposed a model to obtain better estimates of the covariance matrix through shrinkage (Ledoit and Wolf, 2003). Shrinkage reduces estimation error by reducing the sample covariance matrix towards a highly structured estimator, which helps ensure that estimates of the covariance matrix are stable when the number of columns is greater than the number of rows.

While these models improve estimation of the MPT parameters, and therefore provide more accurate tangency portfolios, they still have limitations. Mankert [9] explains that, in practice, the Black–Litterman model can be difficult to implement because it requires that the investor input a large number of estimates including matrices and vectors. Covariance matrix shrinkage and DCC MV-GARCH may provide better estimates of covariance matrices, but are unable to provide estimates of the return vector, which is the other parameter required for mean–variance estimation. Finally, the complexity of the Black–Litterman model and DCC MV-GARCH relative to the Markowitz model makes them difficult for investors to interpret and trust.

We provide a novel approach to relax this assumption. While multiple researchers have attempted to relax this assumption to provide more accurate tangency portfolios by providing better parameter estimates, we propose an alternative approach: forecasting the tangency portfolio through the dimensionality reduction of the efficient frontier's square root second-order polynomial coefficients and selecting the portfolio that is the minimum Euclidean distance from this forecasted tangency portfolio.

This novel approach is related to two previous works. The standard approach to creating the efficient frontier is to use mean–variance optimization to find several efficient portfolios and extrapolate between these, but Merton derived a closed-form solution of the efficient frontier as a square root second-order polynomial function with three coefficients [10]. Alexander et al. [11] previously developed a model to forecast these coefficients to provide better parameter estimates. However, this previous method was more prone to overfit than our proposed method, since the coefficients were calculated using yearly data rather than rolling windows, and used a less-robust weights extraction method that did not use the forecasted tangency portfolio. In addition, the previous paper forecasted the same coefficients that Merton derived, rather than a set of interpretable coefficients. The set of assets used in the previous empirical analysis was also not as diverse as the two sets used in this paper.

This paper provides three main contributions: (1) This paper develops three novel, interpretable coefficients: $r_{MVP}$, $\sigma_{MVP}$, and $u$. (2) This paper provides a method to forecast the coefficients with time-series regression. (3) With these forecasted coefficients, we propose a method of selecting the portfolio that is the minimum Euclidean distance from the forecasted tangency portfolio, because it is the most robust to market movement. We provide empirical results demonstrating that this novel approach outperforms four benchmark portfolios across metrics like Sharpe ratio and max drawdown relative to annual return. We also provide a visual showing that this novel approach consistently outperforms the benchmarks over time (Figure 2 and 3).

*1.1. Assets Data*

To test our proposed models, we used two asset universes: one set corresponding to market capitalization and growth/value, and the other set corresponding to the market sectors. The first set included six mutual funds representing the French–Fama three-factor model's [12] classification of portfolios: growth/value stocks and large/small market capitalization (we refer to this set as GVMC).

The mutual funds in the first set were:
- Fidelity OTC Portfolio—large-cap growth (FDGRX)
- Fidelity Growth and Income Portfolio—large-cap value (FGRIX)
- Fidelity Growth Company Fund—mid-cap growth (FLPSX)
- Fidelity Low-Priced Stock Fund—mid-cap value (FOCPX)
- Invesco Oppenheimer Discovery Fund—small-cap growth (HRTVX)
- Heartland Value Fund Investor class—small cap value (OPOCX)

The second set included all nine original S&P Sector ETFs:
- Materials (XLB)
- Energy (XLE)
- Finance (XLF)
- Industrial (XLI)
- Technology (XLK)
- Consumer Staples (XLP)
- Utilities (XLU)
- Health Care (XLV)
- Consumer Discretionary (XLY)

Both also included a mutual fund to represent an investment in bonds: FPA New Income Fund (FPNIX). These asset datasets were collected using the Yahoo Finance API. In addition, we used the French–Fama three-factor data from Dr. French's website [13] for the risk-free rate of our tangency portfolios.

*1.2. Formulation of Markowitz Efficient Frontiers*

The MPT model performs mean–variance optimization to select portfolios and creates an efficient frontier as a visual. The following is the formal notation used in MPT: $r_i$ is the expected ln return of an asset in $\mathbf{r}$. $w_i$ is the weight of an asset in $\mathbf{w}$, satisfying $\sum_{i=1}^{n} w_i = 1$. $r_{target}$ is the target expected return of the portfolio. $\mathbf{V}$ is the covariance matrix. $\mathbf{e}$ is the ones vector. The formulation of the mean–variance optimization is:

$$\min_{\mathbf{w}} \mathbf{w}^T \mathbf{V} \mathbf{w}$$
$$\text{s.t. } \mathbf{w}^T \mathbf{r} = r_{target} \text{ and } \mathbf{e}^T \mathbf{w} = 1$$

By adding a constraint that $w_i \geq 0$, we can forbid shorting; this paper uses models that allow shorting to better simulate a breadth of investment strategies.

We used both the CVXPY optimizer and the Sequential Least-Squares Quadratic Programming algorithm (SLSQP) in the SciPy package of Python to perform mean–variance optimization for the 33 years of assets data.

*1.3. The Tangency Portfolio*

In MPT, there is often a question of which portfolio to select on the efficient frontier, i.e., what target return to select. Tobin's Separation Theorem [14] states that rational investors would choose a combination of the risk-free asset and an optimal portfolio known as the tangency portfolio. The tangency portfolio is the portfolio that intersects the Capital Market Line (CML) under the Capital Asset Pricing Model (CAPM). The tangency portfolio maximizes a standard portfolio metric, the Sharpe ratio [15], which is defined by $\frac{r_p - r_f}{\sigma_p}$.

The standard approach to finding the tangency portfolio [16] poses it as an optimization problem. Further below, we provide a method to find the tangency portfolio from efficient frontier coefficients.

*1.4. Efficient Frontiers Coefficients*

Merton [10] used Lagrange multipliers to derive a functional representation of the efficient frontier as a square root second-order polynomial with three coefficients [17]:

$$
\begin{aligned}
A &= \boldsymbol{e}^T \boldsymbol{V}^{-1} \boldsymbol{e} > 0 \\
B &= \boldsymbol{r}^T \boldsymbol{V}^{-1} \boldsymbol{e} \\
C &= \boldsymbol{r}^T \boldsymbol{V}^{-1} \boldsymbol{r} > 0
\end{aligned}
\qquad (1)
$$

The equation for the efficient frontier is

$$\sigma^2(r) = \frac{Ar^2 - 2Br + C}{AC - B^2} \qquad (2)$$

## 2. A Novel Set of Interpretable Efficient Frontier Coefficients

We derive an equation with the same form as Equation (2), but with more interpretable coefficients, as

$$\sigma^2(r) = (u^{-1}(r - r_{MVP}))^2 + \sigma_{MVP}^2 \qquad (3)$$

$r_{MVP}$ and $\sigma_{MVP}$ are the return and standard deviation associated with the minimum variance point (MVP). $u$ is the rate of curvature of the efficient frontier utility function, and represents the usefulness of the set of assets returns for mean–variance optimization. An efficient frontier with a higher $u$ diminishes more slowly, and therefore has better tradeoffs of risk and return at every efficient portfolio except for the minimum variance point. We can rearrange terms to calculate each of these more interpretable coefficients as functions of $A, B, C$ as shown in Equation (4).

$$r_{MVP} = \frac{B}{A}, \sigma_{MVP} = \frac{1}{\sqrt{A}}, u = \sqrt{\frac{AC - B^2}{A}} \qquad (4)$$

These more interpretable coefficients each control one graphical efficient frontier component: each coordinate of the vertex, and the rate of curvature. An increase in $r_{MVP}$ implies the market demanding greater returns for all levels of risk. An increase in $\sigma_{MVP}$ implies greater risk for all returns. An increase in $u$ signals a better market for risk-seeking investors.

The $u$ coefficient can be reduced to a more interpretable form in Equation (5).

$$u = \sqrt{\boldsymbol{r}^T \boldsymbol{V}^{-1} \boldsymbol{r}} \cdot \sqrt{1 - S_c(\boldsymbol{r}, \boldsymbol{e})^2} \qquad (5)$$

We can observe that $u$ is a product of the Mahalanobis distance of the return vector to the zero vector and a function of the cosine similarity between the return vector and the vector of ones.

*The Tangency Portfolio Defined by Coefficients*

When using a forecasted efficient frontier, we cannot find the tangency portfolio by solving the standard optimization problem to maximize the Sharpe ratio, because it requires knowledge of the future return vector and covariance matrix. Alternatively, with the efficient frontier coefficients, we can solve for the return and standard deviation of the tangent portfolio. We will start with equation (3). The equation for the line that intersects the efficient frontier and goes through $(\sigma, r)$ is $\sigma - \sigma_P = \sigma'(r)(r - r_P)$, where $\sigma_P$ and $r_P$ are portfolio volatility and return, respectively. The CML must be tangent to the efficient frontier, so we take the derivate of $\sigma(r)$:

$$\sigma'(r) = \frac{r - r_{MVP}}{u^2 \sqrt{(u^{-1}(r - r_{MVP}))^2 + \sigma_{MVP}^2}}$$

The tangent portfolio has an intercept at the risk-free rate, so let $r = r_f$ and $\sigma = 0$, then:

$$0 - \sigma_{TP} = \sigma'(r_{TP})(r_f - r_{TP})$$

$$-\sqrt{(u^{-1}(r_{TP} - r_{MVP}))^2 + \sigma_{MVP}^2} = \frac{r_{TP} - r_{MVP}}{u^2 \sqrt{(u^{-1}(r_{TP} - r_{MVP}))^2 + \sigma_{MVP}^2}} (r_f - r_{TP})$$

Solving for return of the tangency portfolio, $r_{TP}$, yields

$$r_{TP} = \frac{r_{MVP}^2 + u^2 \sigma_{MVP}^2 - r_{MVP} r_f}{r_{MVP} - r_f} \text{ and } \sigma_{TP} = \sigma(r_{TP}) \tag{6}$$

Therefore, to invest in the tangent portfolio of a forecasted efficient frontier, we extract the weights on the efficient frontier at this return.

## 3. Efficient Frontier Forecasting

The model uses an online vector autoregression with exogenous variables (VARX) with lag order 1 to forecast the average efficient frontier coefficients for the forward time period that is the same length as the lookback period used in the mean–variance optimization. In 1980, Sims [18] proposed the use of vector autoregressions in macroeconomic forecasting as a theory-free model, which is still a particularly relevant model to this day. To demonstrate our approach, we elected to limit our feature set to only the lag of the coefficients, and the historical equal-weighted return moving average, as a smaller feature set is less likely to overfit. For each model, we only selected up to two predictors to create more parsimonious models that have greater interpretability. This regularization was incorporated into the model by forcing coefficient parameters of the VARX model in vector notation to 0 as the associating features were not significantly predictive in-sample.

We selected a VARX model to forecast the coefficients so that the results would be interpretable while being able to model autocorrelation. We forecasted the 1 month (21 business days) forward coefficients. The forecasts are online, so after each daily rolling forecast, the current day's coefficients are observed and then included. We elected to use online models as static models do not update as new information arrives. Table 1 shows the out-of-sample R²s for the three proposed online VARX models. The subscript denotes the time the variable is measured, and the superscript denotes the window length used for the calculation of the rolling variable. The predictors all used a 1-year (252 business days) window, demonstrating that long-term efficient frontier coefficients can provide predictive power to forecast the shorter-term future efficient frontier coefficients.

**Table 1.** Out-of-Sample R²s of the online VARX Efficient Frontier Coefficients forecasting models.

| VARX Model | GVMC OoS R² (%) | Sectors OoS R² (%) |
|---|---|---|
| $r_{MVP_{t+21}}^{(21)} = \beta_{1,t} r_{MVP_t}^{(252)} + \beta_{0,t}$ | 13 | 2 |
| $\sigma_{MVP_{t+21}}^{(21)} = \beta_{1,t} \sigma_{MVP_t}^{(252)} + \beta_{0,t}$ | 34 | 9 |
| $u_{t+21}^{(21)} = \beta_{2,t}\sigma_{MVP_t}^{(252)} + \beta_{1,t}\bar{r}_{EW_t}^{(252)} + \beta_{0,t}$ | 1 | 2 |

While these R²s are low, these are measured out-of-sample to ensure the model is not overfitted, and measuring R² out-of-sample is significantly harsher than standard R².

## 4. The Minimum Distance Portfolio to the Forecasted Tangency Portfolio

With the forecasted coefficients, we can find the forecasted tangency portfolio using Equation (6). Because a forecasted efficient frontier does not provide its return vector and covariance matrix, we cannot directly solve for the forecasted tangency portfolio weights. In addition, the forecasted tangency portfolio likely does not exist on the current efficient frontier given the nonstationary behavior of the market. Instead, we are able to solve for the portfolio on the current efficient frontier that is the minimum Euclidean distance from the current efficient frontier to the forecasted tangency portfolio, formulated as

$$\min_r \sqrt{(\hat{r}_{TP} - r)^2 + \left(\hat{\sigma}_{TP} - \sqrt{(u^{-1}(r - r_{MVP}))^2 + \sigma_{MVP}^2}\right)^2}$$

To solve this optimization problem, we can solve for the square of this objective function because the distance function is convex and cannot be negative. This optimization problem can be solved by taking the derivative of the square of the objective function, and solving for the root using Newton's Method.

$$0 = 2(\hat{r}_{TP} - r) + \frac{2(r - r_{MVP})(\hat{\sigma}_{TP}u - \sqrt{(r - r_{MVP})^2 + \sigma_{MVP}^2 u^2})}{u^2\sqrt{(r - r_{MVP})^2 + \sigma_{MVP}^2 u^2}} \quad (7)$$

Now, with the return of the minimum distance portfolio to the forecasted tangency portfolio, we can find the weights by performing mean–variance optimization at this return. Figure 1 provides a visual of the minimum distance portfolio.

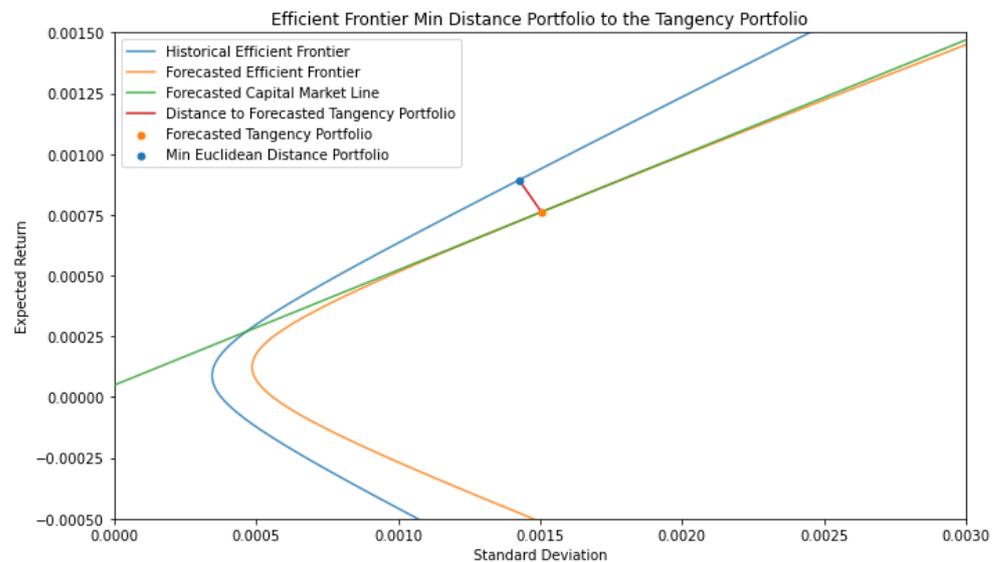

**Figure 1.** The Minimum Distance Portfolio to the Forecasted Tangency Portfolio on 1 February 2018 with a 1-month lookback.

## 5. Empirical Analysis

### 5.1. Benchmarks and Ensuring Realistic Portfolios

To select the best benchmark for our proposed model, we must determine the best version of the tangency portfolio model. The standard tangency portfolio model is sensitive to the update frequency and the lookback period of the historical data. More frequent updates allow the model to use the most relevant data, so we used daily updates. A shorter lookback period uses only the most relevant data, so we used a 1-month window.

To ensure the portfolios do not transfer all of their risk to leverage, we limited the leverage to 1.5× with the following scaling:

$$\widetilde{w}_{<0} = w_{<0} \frac{\frac{l-1}{2}}{\sum(|w_{<0}|)} \text{ and } \widetilde{w}_{>0} = w_{>0} \frac{\frac{l-1}{2}+1}{\sum(|w_{>0}|)}$$

where $l$ is the maximum leverage allowed, and in our case, $l = 1.5$. While this leverage restriction could have been added as a constraint to the optimization, we chose not to do so, so that our empirical results can be generalized to standard tangency portfolios.

Finally, we added 1% daily transaction costs calculated by subtracting the 1% of the sum of absolute changes in weights.

### 5.2. Empirical Results and Discussion

For the first set of assets, the online VARX was initially trained on 1990–1999 data, and the portfolio returns were measured out-of-sample 2000–2022. The second set was trained on 1999–2007 data, and the portfolio returns were measured out-of-sample 2008–2022. We measured the performance of our proposed model against four benchmarks: The Tangency Portfolio with a 1-month rolling window, the equal-weighed portfolio, the S&P 500 total return, and the 60/40 stock and bonds portfolio (60% S&P 500 and 40% FPNIX). Figures 2 and 3 show that the proposed model (black line) outperforms all the benchmarks in terms of return and volatility. Across all years, the proposed model is able to provide consistent returns, while suffering low drawdown during crashes, like in 2008 and 2020. The proposed model in Figure 3 is 2×-levered to provide a better visual comparison, as the proposed model had higher Sharpe than the benchmarks, but lower return.

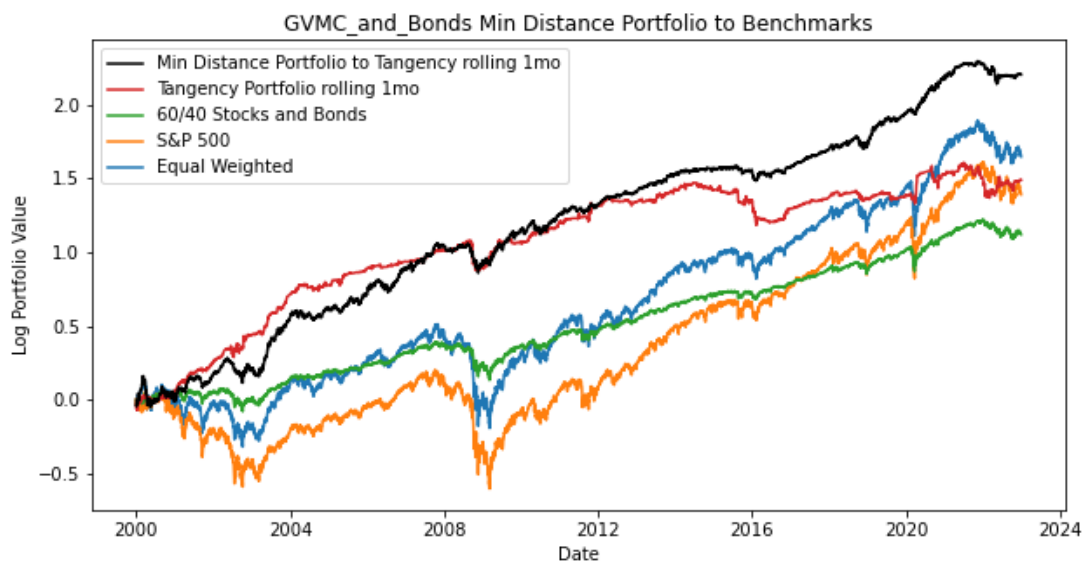

**Figure 2.** The performance of the proposed portfolio model as compared to four benchmarks with the GVMC data.

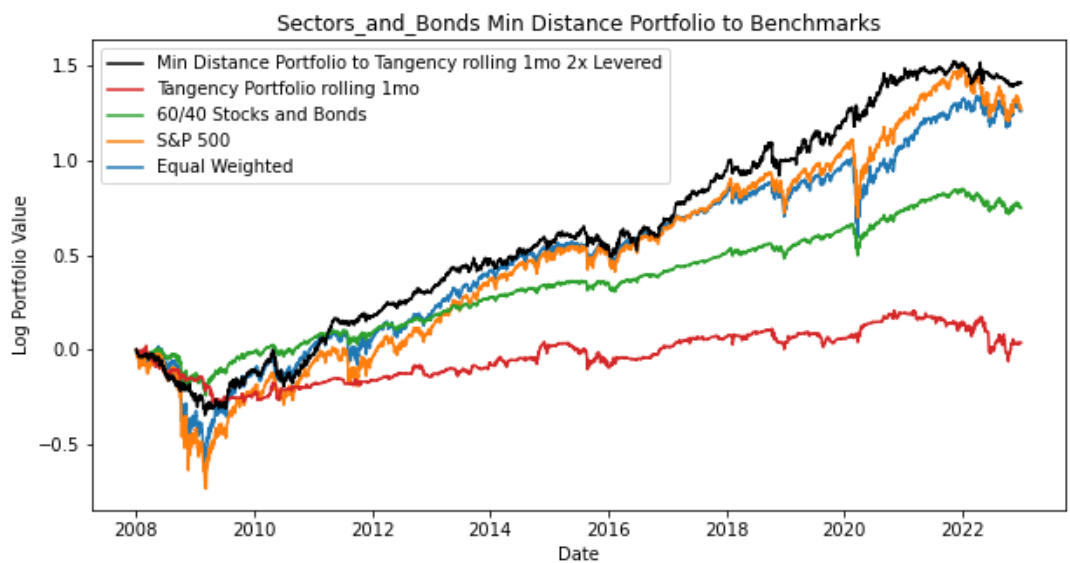

**Figure 3.** The performance of the proposed portfolio model as compared to four benchmarks with the S&P Sector ETFs data.

The proposed model has a greater Sharpe ratio compared to all of the benchmark portfolios, as shown in Tables 2 and 3, for each set of assets, with high annual return relative to max drawdown.

**Table 2.** Portfolio metrics for the market cap and growth/value assets from 2000 to 2022.

| Portfolio | Sharpe Ratio | Sortino Ratio | Annual Return | Max Drawdown |
|---|---|---|---|---|
| Minimum Distance Portfolio to Tangency rolling 1 mo | 1.00 | 1.31 | 10.7% | −18.7% |
| Tangency Portfolio rolling 1 mo | 0.67 | 0.71 | 7.0% | −25.0% |
| Equal Weighted | 0.41 | 0.53 | 9.2% | −50.6% |
| S&P 500 Total Return | 0.33 | 0.42 | 8.3% | −55.3% |
| 60/40 Stocks and Bonds | 0.47 | 0.60 | 5.3% | −22.6% |

**Table 3.** Portfolio metrics for the S&P sector assets from 2008 to 2022.

| Portfolio | Sharpe Ratio | Sortino Ratio | Annual Return | Max Drawdown |
|---|---|---|---|---|
| Minimum Distance Portfolio to Tangency rolling 1 mo 2× Levered | 0.76 | 0.99 | 10.7% | −29.3% |
| Tangency Portfolio rolling 1 mo | 0.01 | 0.01 | 6.3% | −25.8% |
| Equal Weighted | 0.52 | 0.63 | 10.6% | −46.7% |
| S&P 500 Total Return | 0.48 | 0.58 | 11.2% | −51.8% |
| 60/40 Stocks and Bonds | 0.57 | 0.69 | 5.5% | −22.0% |

We conducted alpha regression to demonstrate that the proposed model statistically significantly outperforms the benchmarks, as shown in Table 4.

Table 4. Alpha regressions against baseline portfolios for each universe.

| Universe | GVMC and Bonds | | Sectors and Bonds | |
| --- | --- | --- | --- | --- |
| Baseline Portfolio | Alpha | *p*-Value | Alpha | *p*-Value |
| Tangency Portfolio rolling 1 mo | 0.06 | $<1 \times 10^{-4}$ | 0.10 | 0.0003 |
| Equal Weighted | 0.07 | $<1 \times 10^{-4}$ | 0.06 | 0.01 |
| S&P 500 | 0.06 | $<1 \times 10^{-4}$ | 0.06 | 0.02 |
| 60/40 Stocks and Bonds | 0.06 | $<1 \times 10^{-4}$ | 0.05 | 0.02 |

## 6. Conclusions and Future Research

This paper presents a novel approach to improve out-of-sample Sharpe ratios in comparison to investing in the tangency portfolio. This method is interpretable and does not require input estimates like the Black–Litterman model. This approach decomposes efficient frontiers into three interpretable coefficients: $r_{MVP}$, $\sigma_{MVP}$, and $u$. The utility coefficient, $u$, controls the rate of curvature of the efficient frontier. $u$ measures the value of using the set of assets in mean–variance optimization, as we show that it can be decomposed into the Mahalanobis distance of the return vector to the zero vector and a function of the cosine similarity between the return vector and the ones vector. This method forecasts these coefficients using an online VARX, and determines a forecasted tangency portfolio. This approach invests in the portfolio on the current efficient frontier that is the minimum Euclidean distance from the forecasted tangency portfolio. Using this method out-of-sample from 2000–2022 and 2008–2022 for the two universes, this model outperformed four benchmark portfolios: the tangency portfolio, the equal-weighted, the S&P 500 total return, and the 60/40. The proposed model achieved a higher Sharpe ratio compared to these benchmarks.

While this analysis yielded significant results, there exist certain limitations that we will investigate in future research. The results may have been dependent on the selected assets, so we will test the model on other universes that span the market than the two we tested. The results may also be dependent on the up to 1.5× leverage allowed, so we will perform further analysis on the portfolio risk of using leverage. Another limitation is that the forecasting model used a VARX with only one or two features, so for future research we will explore whether technical indicators such as stochastic oscillators or economic data can improve the regression. Additionally, with a larger feature set, we will employ automated approaches such as Least Absolute Shrinkage and Selection Operator (LASSO) and nonlinear models including decision trees.

**Author Contributions:** Conceptualization, N.A. and W.S.; methodology, N.A. and W.S.; software, N.A.; validation, N.A., and W.S.; formal analysis, N.A. and W.S.; investigation, N.A. and W.S.; resources, N.A. and W.S.; data curation, N.A.; writing—original draft preparation, N.A.; writing—review and editing, W.S.; visualization, N.A.; supervision, W.S.; project administration, W.S.; funding acquisition, W.S. Both authors have read and agreed to the published version of the manuscript."

**Funding:** This research received no external funding

**Institutional Review Board Statement:** Not applicable

**Informed Consent Statement:** Not applicable

**Data Availability Statement:** Publicly available datasets were analyzed in this study. This data can be found on Yahoo Finance here: https://finance.yahoo.com/.

**Conflicts of Interest:** The authors declare no conflict of interest.

## References

1. Markowitz, H. Portfolio selection. *J. Financ.* **1952**, *7*, 77. https://doi.org/10.2307/2975974.
2. Sharpe, W.F. Capital asset prices: A theory of market equilibrium under conditions of risk. *J. Financ.* **1964**, *19*, 425–442. https://doi.org/10.1111/j.1540-6261.1964.tb02865.x.

3. Davis, M.H.A.; Norman, A.R. Portfolio Selection with Transaction Costs. *Math. Oper. Res.* **1990**, *15*, 676–713. https://doi.org/10.1287/moor.15.4.676.
4. Dickinson, J.P. The Reliability of Estimation Procedures in Portfolio Analysis. *J. Financ. Quant. Anal.* **1974**, *9*, 447–462. https://doi.org/10.2307/2329872.
5. Kan, R.; Zhou, G. Optimal Portfolio Choice with Parameter Uncertainty. *J. Financ. Quant. Anal.* **2007**, *42*, 621–656. https://doi.org/10.1017/S0022109000004129.
6. Black, F.; Litterman, R. Global portfolio optimization. *Financ. Anal. J.* **1992**, *48*, 28–43. https://doi.org/10.2469/faj.v48.n5.28.
7. Engle, R.F.; Sheppard, K. *Theoretical and Empirical Properties of Dynamic Conditional Correlation Multivariate GARCH*; Working Paper No. 8554; National Bureau of Economic Research: Cambridge, MA, USA, 2001. https://doi.org/10.3386/w8554.
8. Ledoit, O.; Wolf, M. Improved estimation of the covariance matrix of stock returns with an application to portfolio selection. *J. Empir. Financ.* **2003**, *10*, 603–621. https://doi.org/10.1016/S0927-5398(03)00007-0.
9. Mankert, C. The Black-Litterman model: Towards its use in practice. Ph.D. Thesis, KTH Royal Institute of Technology, Stockholm, Sweden, 2010. Available online: http://urn.kb.se/resolve?urn=urn:nbn:se:kth:diva-26798 (accessed on 1 November 2019).
10. Merton, R.C. An Analytic Derivation of the Efficient Portfolio Frontier. *J. Financ. Quant. Anal.* **1972**, *7*, 1851.
11. Alexander, N.; Scherer, W.; Burkett, M. Extending the Markowitz model with dimensionality reduction: Forecasting efficient frontiers. In IEEE Xplore, Proceedings of the 2021 Systems and Information Engineering Design Symposium (SIEDS), Charlottesville, VA, 2021, 1–6. https://doi.org/10.1109/SIEDS52267.2021.9483775.
12. Fama, E.F.; French, K.R. Common Risk Factors in the Returns on Stocks and Bonds. In *The Fama Portfolio: Selected Papers of Eugene F. Fama*; University of Chicago Press: Chicago, IL, USA, 1993; pp. 392–449.
13. Kenneth r. French—Data Library. Available online: https://mba.tuck.dartmouth.edu/pages/faculty/ken.french/data_library.html (accessed on 28 March 2023).
14. Tobin, J. Liquidity Preference as Behavior Towards Risk. *Rev. Econ. Stud.* **1958**, *25*, 65. https://doi.org/10.2307/2296205.
15. Sharpe, W.F. Mutual fund performance. *J. Bus.* **1966**, *39*, 119–138. Available online: https://www.jstor.org/stable/2351741 (accessed on 1 November 2019).
16. Luenberger, D.G. *Investment Science*, 1st ed.; Oxford University Press: Oxford, UK, 1997.
17. Petters, A.O.; Dong, X. *An Introduction to Mathematical Finance with Applications: Understanding and Building Financial Intuition*, 1st ed.; Springer: Cham, Switzerland, 2016.
18. Sims, C. Macroeconomics and Reality. *Econometrica* **1980**, *48*, 1–48. https://doi.org/10.2307/1912017.